\documentclass{ametsoc}

\journal{jas}

\bibpunct{(}{)}{;}{a}{}{,}

\title{A Mat\'ern based multivariate Gaussian random process for a consistent model of the horizontal wind components and related variables}

\authors{R\"udiger Hewer\correspondingauthor{Meteorological Institute, University Bonn, Auf dem H\"ugel 20, Bonn, Germany}, Petra Friederichs, Andreas Hense}

\affiliation{Meteorological Institute, University Bonn, Auf dem H\"ugel 20,  Bonn, Germany}

\email{rhewer@uni-bonn.de}

 \extraauthor{Martin Schlather}
    \extraaffil{School of Business Informatics and Mathematics, University Mannheim, B3, Mannheim, Germany}

\newcommand{\R}{\ensuremath{\mathbb{R}}}   
\newcommand{\E}{\ensuremath{\mathbb{E}}}  

\newcommand{\Cov}{\ensuremath{\text{Cov}}}

\newcommand{\defeq}{\mathrel{\mathop:}=}  

\renewcommand{\div}{\nabla\cdot}
\newcommand{\rot}{\nabla \times}

\newcommand\minus{%
 \scalebox{0.75}[1.0]{\( - \)}
}

 \usepackage{amsmath}   
  \usepackage{amsthm}    
  \usepackage{amssymb}
 \usepackage{theoremref} %
\usepackage{amssymb} 

\usepackage{graphicx}
\usepackage{caption}
\usepackage{subcaption} 
\usepackage {lmodern}
\usepackage{dsfont}
\usepackage {graphicx}
\usepackage[fleqn]{nccmath}
\usepackage {ulem}
\usepackage{floatrow}
\usepackage{titleref}
\makeatletter
\newcommand*{\currentname}{\TR@currentTitle}
\makeatother

  \theoremstyle{definition}

    {\begin{proof}[Beweis]}
    {\end{proof}}
  
\renewcommand*{\figurename}{Figure} 
\renewcommand*{\tablename}{Table}

\begin{document}

\maketitle

%








\section {Introduction}

An appropriate representation of the covariance structure in spatial models of  meteorological variables is  essential when analyzing \citep {gandin1963,Kalnay03} meteorological data using data assimilation  \citep{Hollingsworth1986, Evensen94, bonavita2012,pu2016}. This generally requires an appropriate representation of the background error covariance matrix.
Further, spatial stochastic models for meteorological variables should  respect physical relationships. 

One of the first approaches to include physical consistency via differential relations between variables can be found in \citet{Kolmogorov1941}. \cite{Thiebaux1977} introduced a covariance model for wind fields assuming geostrophic balance, thereby incorporating anisotropy in the geopotential height.
\citet{Daley1985} derived a covariance model for the horizontal wind components assuming a Gaussian covariance model for the velocity potential and the streamfunction, where he derived the differential relations between the potentials and the wind field. The covariance model  proposed by \citet{Daley1985} is rather flexible as it allows for geostrophic coupling,  non-zero correlation of streamfunction and velocity potential, and differing scales for the two potentials.  \citet{Daley1985} also considered geopotential height as an additional model variable. However, the resulting covariance function for the wind fields is not positive definite for many parameter combinations. 
\cite{Hollingsworth1986} adapted Daley's method and formulated a covariance function for the potentials using cylindrical harmonics. They show that on the synoptic scale the correlation between the potentials is small, such that \cite{Daley1991} reformulated his model for zero correlations.
These approaches \citep {Thiebaux1977,Hollingsworth1986,Daley1985} as well as our model differ from current data assimilation methods, as they provide an explicit, parametric and analytic covariance model for the background error. So-called \textit {control variable transform} methods \citep {bannister2008} describe the background error matrix in an implicit non-parametric way via its square root \footnote {e.g. Cholesky decomposition} using latent variables which model the physical variables. Sample  based methods like the ensemble Kalman filter \citep {Evensen94} describe the error statistics based on estimates obtained from an ensemble.

The data assimilation literature \citep[e.g.][] {Thiebaux1977,Hollingsworth1986,Daley1985} typically uses the stochastic models in order to describe the covariance matrix of the background error, which is the difference of the a forecast and the true field. Similar methods have also been used in order to describe the full turbulent field \citep {Frehlich2001}. There has also been considerable interest in describing the statistics of the velocity field directly or via its spectrum \citep {buhler2014,lindborg2015,bierdel2016}.

While \cite {Thiebaux1977}, \cite {Hollingsworth1986}, and \cite{Daley1985}  include physical relations via differentiation of the  covariance function,
finite difference operators are used in Bayesian hierarchical models.  For example, \cite{Royle1999}  modeled the geostrophic relation of pressure and wind field.

In this paper, we propose a multivariate Gaussian random field (GRF) formulation for six atmospheric variables in a horizontal two-dimensional Cartesian space. Assuming a bivariate Mat\'ern covariance for streamfunction $\psi$ and velocity potential $\chi$, we derive the covariance structure of  the horizontal wind  components $\vec {U}=\left( u, v\right)^T$ as well as vorticity $\nabla \times \vec {U}\defeq -\frac{\partial}{\partial e_2}u+\frac{\partial}{\partial e_1}v$ and divergence $\nabla \cdot \vec {U}$. All of these quantities are connected via the Helmholtz decomposition, which states that for any given wind field $\vec {U}$ there exists a streamfunction $\psi$ and velocity potential $\chi$, such that
$\vec {U}= \nabla \times \psi +\nabla \chi$,
where $\nabla \times \psi\defeq \big(-\frac{\partial}{\partial e_2}\psi, \frac{\partial}{\partial e_1}\psi \big)^T$. 
In dimension two and with appropriate boundary conditions this decomposition is unique. Curl and divergence of the wind field are given as $\nabla \times \vec {U}=\Delta \psi$ and $\div \vec {U}=\Delta \chi$, respectively,
 where $\Delta$ is the  2-dimensional Laplace operator. 

Our multivariate GRF formulation is novel for several reasons. While e.g.\ \citet{Daley1985} only used the potentials to derive the covariance function of the wind fields, our model is formulated for all related variables, including a formulation for the potential functions and the wind field, as well as vorticity and divergence. 
Secondly, our model provides a formulation for anisotropy in the wind field and the related potentials. 
Further, we allow for non-zero correlations between the rotational and divergent wind component, which might be particularly relevant for  atmospheric fields on  sub-geostrophic scales. 
We show that the scale parameters considered by \cite {Daley1985} are inconsistent with non-zero correlations between streamfunction and velocity potential, as they do not lead to a positive definite model. An exact derivation of the condition under which the covariance function of Daley's model is positive definite is given in the appendix. Further our model is a counter example to a theorem of \cite {obukhov1954statistical}, which claims that there is no isotropic wind field with non-zero correlation of the rotational  and non-rotational component of the wind field. More details to Obukhovs claim are given in the appendix.

The covariance function of our multivariate GRF will be incorporated into an upcoming version of the spatial statistics R package RandomFields \citep{RandomFields}. This  opens the possibility for a wealth of applications  in spatial statistics, including the conditional simulation of streamfunction and vector potential given an observed wind field, a consistent formulation of the covariance structure for both the potential and the horizontal wind components to be used in data assimilation, or stochastic interpolation (kriging) of each of the involved variables given the others. Kriging is the process of computing the conditional expectation of a certain variable given others. It is typically used to interpolate fields.

To exemplify the multivariate GRF we estimated its parameters for atmospheric fields of the numerical ensemble weather prediction system, COSMO-DE-EPS \citep{Gebhardt11}, provided by the German Meteorological Service (DWD). COSMO-DE is a high-resolution forecast system, that provides forecasts on the atmospheric mesoscale \citep{Baldauf11}. Estimation is realized using the maximum likelihood method, while uncertainty in the parameter estimation is assessed by  parametric bootstrap \citep{Tibshirani1994}. We also discuss the meteorological relevance of the parameters.

The remainder of the paper is organized as follows. In Section \ref {Theory} we introduce the multivariate GRF, and demonstrate how the physical relations and anisotropy are included in the model formulation. Section \ref {Data} introduces the COSMO-DE-EPS data.  Section \ref {Methods} is devoted to the parameter estimation and the assessment of the uncertainties, while Section \ref {Results} presents and interprets the results of the estimation. We conclude in Section \ref {Conclusions}  and discuss potential applications, limitsand extensions of our multivariate GRF.

\section {Theory}
\label {Theory} 
An important aspect of our multivariate GRF is the inclusion of the differential relations between the atmospheric variables.
Under  weak regularity assumptions the derivative of a Gaussian process is again a Gaussian process \citep{Adler2007}.  Hence, the assumption of Gaussianity of the streamfunction and the velocity potential implies Gaussianity of all  the considered variables.  A zero-mean Gaussian process is uniquely characterized by the  covariance function, we only need to study the joint covariance of a random field and its derivatives. A Gaussian process  $\left( X_s,s\in \R^d\right)$ is a continuously indexed stochastic process. For each finite number of locations $\left( s_i, i=1,\dots ,n\right)$ the variables $\left( X_{s_i},i=1,\dots,n\right)$ have a multivariate Gaussian distribution.

Let $ X_s,s\in \R$, be a stochastic process with finite second moments, and assume that the covariance function $C(s,t)=\Cov \left( X_s,X_t\right)$ is twice continuously differentiable, then the covariance model of the process and its mean-square derivative is given by
\begin{align} 
 \Cov \left( \left(\begin{array}{c} X_s \\ d_s X_s \end{array}\right),\left(\begin{array}{c} X_t \\ d_t X_t \end{array}\right)\right)=\left(\begin{matrix}
\Cov \left( X_s,X_t\right) & d_t\Cov \left( X_s,X_t\right)  \\
d_s \Cov \left( X_s,X_t\right) & d_sd_t\Cov \left( X_s,X_t\right)
\end{matrix}\right),\label {cov_deriv}
 \end{align}
 where $s,t\in \R$ \citep {ritter2000average}.
 Using the linearity in the arguments the validity of this equation can be roughly seen by
 \begin{eqnarray*} 
\Cov \left( X_s,d_tX_t\right)&=&\lim_{\Delta\to 0} \Cov \left( X_s,\frac{X_t-X_{t+\Delta}}{\Delta}\right)\\
 &=&\lim_{\Delta \to 0}\frac{\Cov \left( X_s,X_t\right)-\Cov \left( X_s,X_{t+\Delta}\right)}{\Delta}\\
 &=&d_t\Cov \left( X_s,X_t\right).
 \end{eqnarray*}
One key advantage of this approach is that the bivariate covariance in  $\eqref{cov_deriv}$  allows us to model the dependence between the process and its derivative. In order to provide a better theoretical basis for this idea, we consider the following definiton.
\newtheorem* {Definition} {Definition}
\begin {Definition} 
A stochastic process $X_t,t\in \R^d$, is mean square differentiable at $t\in \R^d$ in direction $e_i,  i=1,\dots,d$, if there exists a random variable $X_t^{\left( i\right)}$ with $\E\left(X_t^{\left( i\right)}\right)^2< \infty $ such that,
\begin{align*} 
 \E \left( \left( \frac{X_t-X_{t+\Delta e_i}}{\Delta}\right)-X_t^{\left( i\right)}\right)^2\to 0 \quad \mbox{ as } \quad  \Delta \to 0,
 \end{align*}
where $e_i$ denotes the unit vector in the $i-th$ coordinate direction.
 In this case, we use the following notation $\frac{\partial}{\partial e_i}X_t=X_{t}^{\left( i\right)}$. 
 \end {Definition}
A stochastic process is mean square differentiable if its covariance function is twice continuously differentiable \citep{ritter2000average}. However, this condition is neither sufficient nor necessary for the differentiability of the sample paths. For Gaussian processes the following conditions on the derivatives of the process guarantees continuity of the sample paths.
The paths of a Gaussian process are continuous, if there exist $0<C<\infty$ and $\alpha,\eta>0$ such that
\begin{align*} 
\E \left| \frac{\partial}{\partial s}X_s-\frac{\partial}{\partial t}X_t\right|^2\le \frac{C}{\left|\log \left|s-t\right|\right|^{1+\alpha}},
 \end{align*}
for all $\left|s-t\right|<\eta$, see Theorem 1.4.1. in \cite{Adler2007}.

In our case, the covariance function describes the dependence of the horizontal wind components $u_s$ and $v_s$, streamfunction $\psi$, velocity potential $\chi$,  and the Laplacian of the potentials (i.e.\ vorticity $\zeta = \Delta \psi$ and divergence $D = \Delta \chi$) at locations $s, t \in \R^2,$
\begin{align}
\label {full_covariance}
  &C(s,t)=\Cov \Big( \big(\begin{array}{c}\!\psi_{s},\ \chi_{s},\ u_s ,\ v_s ,\ \Delta \psi_{s},\ \Delta \chi_{s}\!\end{array}\big)^T,\big(\!\begin{array}{c}\!\ \psi_{t},\ \chi_{t},\ u_t ,\ v_t,\ \Delta \psi_{t},\ \Delta \chi_{t}\!\end{array}\big)^T\Big) .
\end{align}

The covariance function $C(s,t) $ is well-defined, if   
\begin{align*} 
 C _{\psi,\chi}\left( s,t\right)= \Cov \left( \left(\begin{array}{c} \psi_s \ \chi_s \end{array}\right)^T,\left(\begin{array}{c} \psi_t  \ \chi_t \end{array}\right)^T\right) 
\end{align*}
 is four times continuously differentiable. Four times differentiability of the covariance function is equivalent to the process being twice mean square differentiable, see Lemma 14 in \cite{ritter2000average}.  
 
In the remainder of the paper we will consider stationary processes, which means that $C(s,t)$ depends only on the lag vector $h=t-s$. We will adopt a commonly used notation for stationary processes, $C (h)\defeq C \left( 0,h\right)$.
Our next step is to review two notions of isotropy that exist for multivariate processes. Following \cite{Schlather2015} a vector of scalar quantities is called isotropic if the covariance function $C$ fulfills
\begin{align} 
 C \left( Qh\right)=C \left( h\right) \quad \ h\in\R^d,
 \label {isotropy}
 \end{align}
 for all rotation matrices $Q$ and $h=t-s$. A matrix Q is a rotation matrix if $QQ^T$ equals the d-dimensional identity matrix and $\det(Q)=1$. Under the assumption of stationarity \eqref {isotropy} is equivalent to the more typically used notion of isotropy $C \left( h\right)=C \left( \|h\|\right)$. Bi- (multi-) variate variables consisting of scalar quantities such as streamfunction, velocity potential or the Laplacian thereof fulfill \eqref {isotropy}.
 A multivariate process is  \textit{vector isotropic} if its covariance functions fulfills
 \begin{align} 
 C \left(h\right)=Q^TC(Qh)Q \quad \text {for all}\ h\in\R^d.
 \label {vector_isotropy}
 \end{align}
This relation shows that $\E \left( X_0X_h^T\right)=\E \left( Q^T X_0 \left(Q^TX_{Qh}\right)^T\right),$ which means that the covariance is preserved if the lag vector $h$ and the random vector are rotated simultaneously.

In the remainder of the paper we consider isotropic processes, hence $C_{\psi,\chi}\left( Qh\right)=C_{\psi,\chi}\left( h\right)$ for all rotation matrices $Q$.
Using the notation,
\begin{align}	A=\left(\begin{matrix}
r_1 \cos\theta & r_1 \sin \theta  \\
-r_2 \sin  \theta & r_2 \cos  \theta
\end{matrix}\right),
\label {anisotropy_transform}
\end{align}
we set $C_{\psi,\chi,A}\left( h\right)= C_{\psi,\chi}\left( Ah\right)$.

The effect of the anisotropy matrix $A$ on the covariance function of the vector components, namely the rotational part $\nabla \times \psi$ and the divergent part $\nabla \chi$, is non-trivial. The divergent part satisfies
\begin {align} 
\label {divergent}
 \Cov \left( \nabla \chi \left( As\right),\nabla \chi \left( At\right)\right)
=A^T\Cov \left(\left( \nabla \chi\right) \left( As\right),\left( \nabla \chi\right)\left( At\right)\right)A.
 \end {align}
The rotational part fulfills a more complex formula
\begin {align}
\label {rotational}
& \Cov \left( \nabla \times \psi \left( As\right),\nabla \times \psi \left( At\right)\right)\\
&=RA^TR^T\Cov \left( \left(\nabla \times \psi\right) \left( As\right),\left( \rot \psi\right)\left( At\right)\right)RAR^T, \notag
 \end {align}
 where 
 \begin {align}	R=\left(\begin{matrix}
0 & -1  \\
1 & 0
\end{matrix}\right).\notag
\end {align}

If $A$ is simply a rotation matrix (i.e. $r_1=r_2=1$), then $RAR^T=A$, which implies that both the divergent and the rotational part are vector-isotropic. 
For the Laplacians we obtain the following transformation
\begin{eqnarray} \label {Laplacian}
\Cov \left( \Delta\chi \left( As\right),\Delta \chi \left( At\right)\right)
 &=& r_1^4 \Cov \left(\left.\partial^2_{e_1}\chi\right|_{As},\left.\partial^2_{e_1}\chi\right|_{At} \right)+r_2^4 \Cov \left( \left.\partial^2_{e_2}\chi\right|_{As},\left.\partial^2_{e_2}\chi\right|_{At}\right)\notag\\
 &+&2r_1^2r_2^2 \Cov \left(\left.\partial^2_{e_1}\chi\right|_{As},\left.\partial^2_{e_2}\chi\right|_{At} \right)
 .
 \end{eqnarray} 
 In the appendix we provide the formulae for all entries of the covariance  matrix \eqref {full_covariance} in the isotropic case.  Equations $\eqref {divergent}\! -\!\eqref {Laplacian}$ are useful since they are the easiest way to compute the covariance in the anisotropic case from the covariance in the isotropic case. They have been derived using the chain rule and the linearity of the covariance function in both arguments.

 Our GRF  is a counter example to a theorem of \cite{obukhov1954statistical}, which claims that the rotational and divergent component of isotropic vector fields are necessarily uncorrelated, which is equivalent to streamfunction and velocity potential being uncorrelated. Obukhov considers an invalid expression for the covariance of a rotational field and deduces from this expression that it is necessarily uncorrelated to a gradient field. We present the detailed argument in the Appendix.

In the remainder of the paper we will exemplify the full process in the case that the potential functions have the following bivariate structure.
 \begin{align} 
  C_{\psi,\chi}\left( s,t\right)=  \left(\begin{matrix}
\sigma_\psi^2 & \rho\sigma_\psi\sigma_\chi  \\
 \rho\sigma_\psi\sigma_\chi & \sigma_\chi^2 
\end{matrix}\right)M\left(\|A\left( t-s\right)\|_2, \nu\right),
\label {matern}
 \end{align}
where $M\left(\cdot, \nu\right)$ denotes the Mat\'ern correlation function with smoothness parameter $\nu$, and $\|t-s\|_2$   the $L^2$ norm.  \cite {Goulard1992} consider a more general model and prove its positive definiteness, implying the positive definiteness of our model \eqref {matern}.

\begin{figure}
   \includegraphics*[width=1\textwidth]{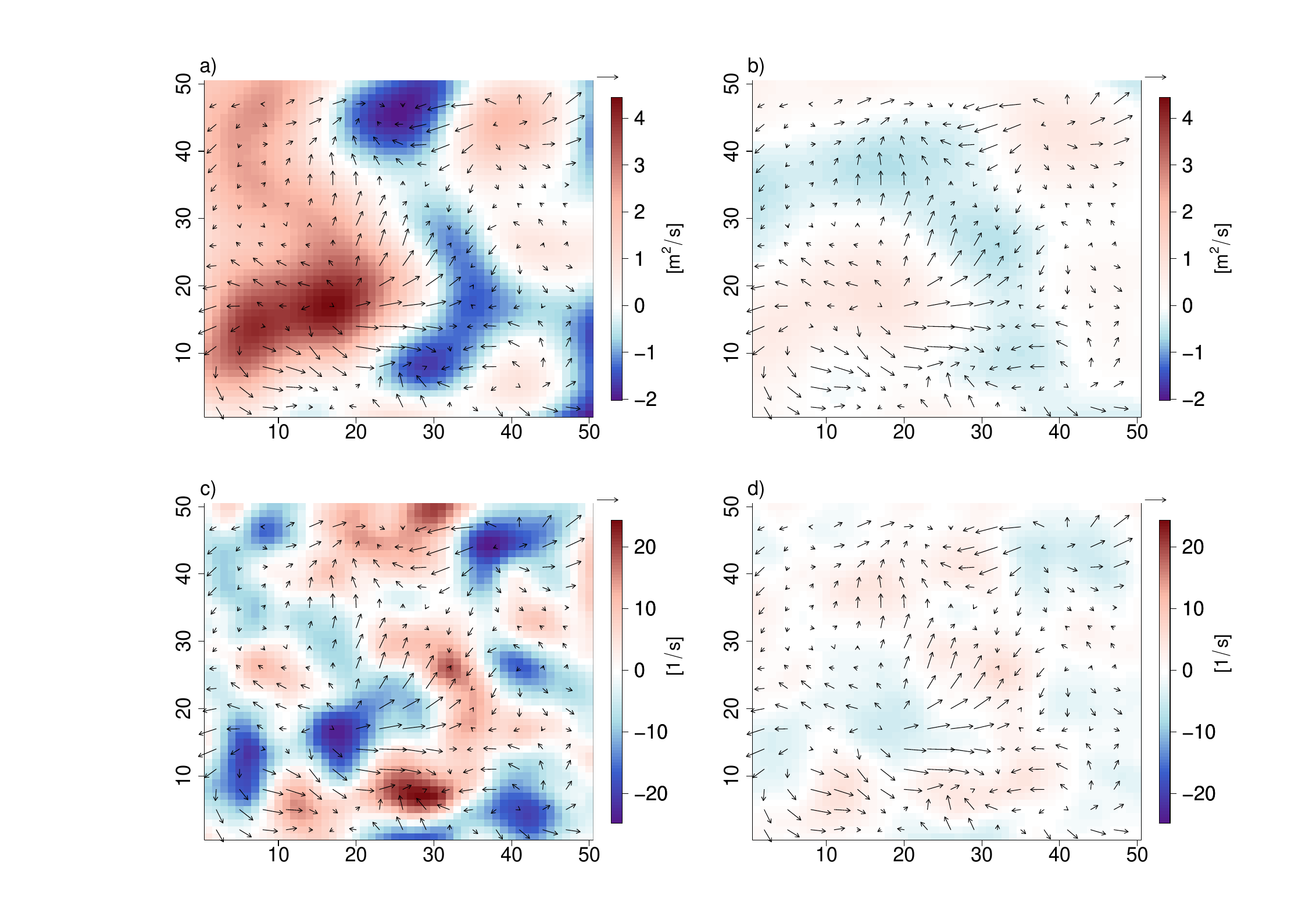}
   \caption {Isotropic realization of the multivariate GRF with parameters  $\nu\!=\!5$,\ $\sigma_\chi/\sigma_\psi\!=\!0.3,\ \rho\!=\!0.7,\ r_1\!=\!r_2\!=\!0.25$. In color are shown a) streamfunction, b) velocity potential, c) vorticity, and d) divergence.  The arrows represent the associated wind fields in m/s. The arrow in the right upper corner is a standard arrow of 0.5 $m/s$. The x/y-axis indicate distance measured in grid points.}
    \label {fig:6_dim_field}
\end{figure}

Fig. \ref{fig:6_dim_field} represents a realization of the full stochastic process, with parameters chosen in order illustrate the flexibility of the model. The rotational wind component is larger than the divergent wind component with a ratio of $\sigma_\chi/\sigma_\psi = 0.3$. The two potential functions are strongly correlated with a correlation coefficient of $\rho = 0.7$.  The coherence of the variables can be very well spotted, although the simulation of the process is inherently stochastic.
The smoothness is set to $\left( \nu=5\right)$, which implies that not only the potentials but also vorticity and divergence are continuously differentiable. We will see later in Section \ref{Methods}, that realistic mesoscale wind fields have a smoothness parameter close to $1.25$. This suggests that the vorticity and divergence fields are dis-continuous.

\section{Data}\label {Data}
The horizontal wind fields are taken from the numerical weather
prediction (NWP) model COSMO-DE, namely the wind fields at model level 20
(i.e.\ at approximately 7 km height).
COSMO-DE is the operational version of the non-hydrostatic limited-area NWP model COSMO (Consortium of Small-scale Modeling) operated by DWD \citep{Baldauf11}. It provides forecasts over Germany and surrounding countries on a 2.8 km horizontal grid and 50 vertical levels. At this grid size deep convection is permitted by the dynamics, and COSMO-DE is able to generate deep convection without an explicit parameterization thereof. Thus COSMO-DE particularly aims at the prediction of mesoscale convective precipitation with a forecast horizon of up to one day. The ensemble prediction system (COSMO-DE-EPS) uses COSMO-DE with different lateral boundary conditions (LBC), perturbed initial conditions, and slightly modified parameterizations. The four LBC are generated by the Global Forecast Systems of NCEP, the Global Model of DWD, the Integrated Forecast System of ECMWF and the Global Spectral Model of the Meteorological Agency of Japan.  For details on the setup of COSMO-DE-EPS the reader is referred to \citet{Gebhardt11}, \citet{Peralta12}, and references therein.

In our application we concentrate on a COSMO-DE forecast for 12 UTC  on 5 June 2011 initialized on 00 UTC. COSMO-DE-EPS provides 20 forecasts of horizontal wind fields on a grid with $461\times421$ grid points. Five ensemble members are forced with identical LBC, respectively. They only differ due to perturbed initial conditions and four different parameterizations. Thus differences between the members with identical LBC are mainly due to small-scale internal dynamics. These differences are the differences obtained from subtracting two fields which have been generated using the same lateral boundary conditions. All combinations of fields with different model physics and identical lateral boundary conditions generate a set of 40  different fields of differences.
The differences are referred to as inner-LBC anomalies.

\begin{figure}
\centering
 \includegraphics[width=1\textwidth]{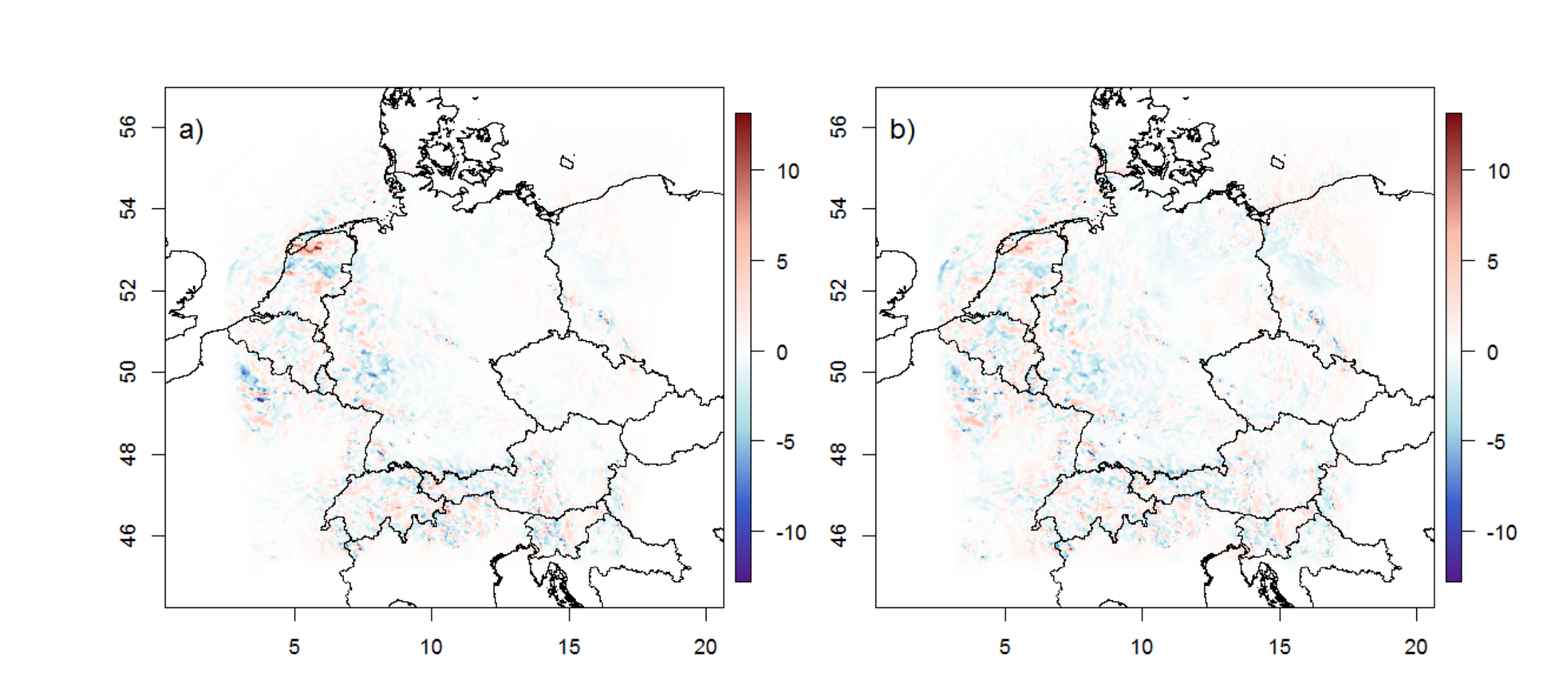}  
  \caption {Zonal wind component at 12 UTC on 5 June 2011. a) Shows the inner-LBC anomalies, b) the transformed inner-LBC anomalies. The colors represent wind speed in m/s. The x/y- axis are in longitude and latitude.} 
    \label {fig:simuset1}
\end{figure}
\begin {figure}
\includegraphics*[width=\textwidth,page=1]{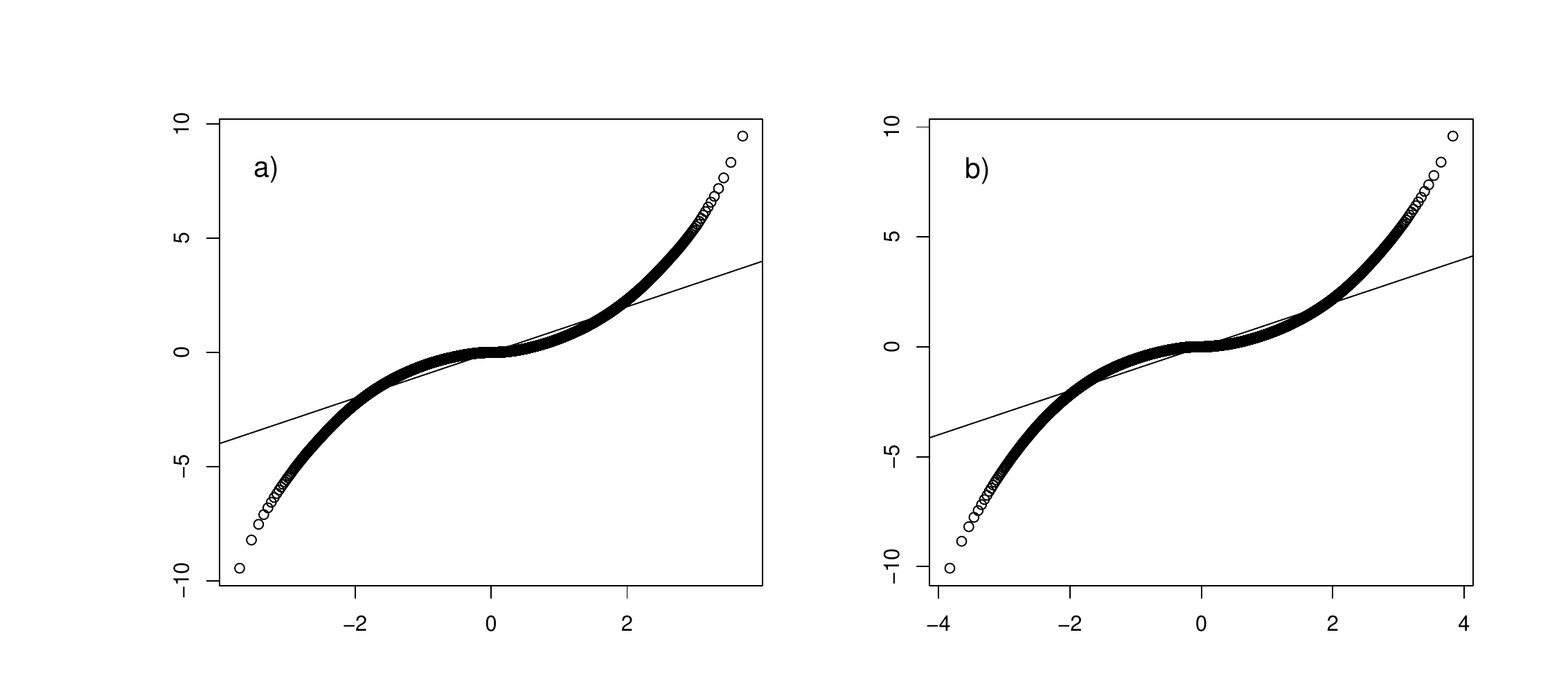}
\caption { Quantile-Quantile plot of  (a) the zonal, and  (b) the meridional wind component of transformed inner LBC anomalies versus a standard normal distribution. The linear lines indicate perfect accordance with the marginal distributions, both graphs depict clear deviation from the normal distribution.}
\label {fig:marginal}
\end {figure}

To illustrate the data, Fig. \ref{fig:simuset1} displays a field of inner-LBC anomalies of the zonal wind component. The fields exhibits small scale anomalies with amplitudes that vary over the model region while the spatial structure seems relatively homogeneous. Thus, the data  violate the assumption of stationarity.
In order to model the instationarity of the variance we estimate the spatial kinetic energy $\widehat g$ by applying a kernel smoother to the kinetic energy field. In analogy to the field of electric susceptibility $\left( 1+\chi_e\right)$ which models the spatial varying potential polarization of the dielectric medium \citep {jackson1962}, we apply the following transformation to the data
\begin{align*} 
 \widetilde U_s=\frac{U_s}{c+\widehat g_s},
 \end {align*}
 where $c\in\R_{+}$.
 Such a transformation, if applied to the full field $\left(\widetilde \chi,\widetilde\psi,\widetilde U,\widetilde D,\widetilde\zeta\right)=\left( \chi,\psi,U,D,\zeta\right)/(c+\widehat g)$, violates the differential relations that hold between the variables, though they are still valid approximately.  For example for a non-rotational field we have
 \begin{align}
 \label {transform} 
 \nabla\left(\frac{ \chi}{c+\widehat g}\right)=\frac{\nabla \chi}{c+\widehat g}+\varepsilon.
 \end {align}
 The smoother the transformation the smaller the approximation error
 \begin{align*} 
  \varepsilon=-\frac{\chi\nabla \left( c+\widehat g\right)}{\left( c+\widehat g\right)^2}.
 \end {align*}
Due to the constant $c>0$ the transformation \eqref {transform} does not resolve the full instationarity of the data. Still we find that this transformation is superior to the more natural transformation $\widetilde U=U/\widehat g$, as the approximation error for the potential functions is strongly reduced by the introduction of $c>0$.	We observe a trade-off between the differential relations being hardly violated and on the other side Gaussian marginal distribution and constant variance in space by a rougher function $\widehat g$ and values of $c$ close to zero. We chose $c=1/3$ and a kernel such that the transformation kurtosis of the data is reduced from 24 to 16, while we have to accept  an error of the potential fields close to 15 percent. The  error is measured by comparing the potential that satisfies $\nabla\widetilde \chi=U/(c+\widehat g) $ and the potential that satisfies $\nabla \chi=U$ and is normalized by $c+\widehat g$ (the same is done  for the rotational part).
Figure \ref {fig:simuset1}  shows that the instationarity of the original fields is mitigated by the transformation.
Figure \ref {fig:marginal} shows

the marginal distribution of the transformed inner-LBC anomalies for the zonal and the meridional wind component. Both distributions deviate from the assumption of Gaussian marginals, although Gaussianity is a common assumption for wind fields in the meteorological literature \citep{Frehlich2001}. The kurtosis amounts to about 16 instead of 3, which results in heavier extreme values than expected under the  assumption of Gaussianity.

\section {Parameter estimation}\label {Methods}

 We start by parameter estimation of the bivariate GRF model for the transformed inner-LBC anomalies of the horizontal wind fields described in Section \ref {Data}. Since the computation of the Gaussian likelihood would require  the inversion of a quadratic matrix with $2\times 461\times421$ rows, a standard maximum likelihood approach is unfeasible.
We thus use a composite likelihood (CL) approach to approximate the true likelihood function. An overview of the CL approach is given in \citet{Varin2011}. Here, we apply a special version of the CL approach known as pairwise likelihood \citep{Cox2004}. For a \textit {bivariate} field this likelihood is a product of 4-dimensional likelihoods. We calculate the log likelihood of the CL as
\begin{align*} 
l^{c} \left( \theta\right)=\sum_{s\in \mathbb {G}}\sum_{h\in N} \text {log}\left(  L\left.\left(u_{s},v_{s}, u_{s+h},v_{s+h} \right\vert \theta \right)\right),
 \end{align*}
 where $\theta$ denotes the parameter vector, and $\mathbb{G}$ denotes the set of all  grid points. The set $N$ controls for which separations $h$ the likelihood is computed. The set $N$ has to be determined relative to the given problem. If feasible it should include all lags $h$ for which there is non-negligible dependence and some for which there is negligible dependence, in order to estimate the range. One way of determining this is to inspect  the empirical covariance estimate. We chose $N$ to be a regular $41\times 41$ grid with step size one, which is centered in the origin. 
 The choice is justified by the low uncertainties observed in the parametric bootstrap samples presented below. 

The unknown parameters are the variances of the potentials $\sigma_\psi^2$ and $\sigma_\chi^2$, their correlation $\rho$, the smoothness parameter $\nu$, and the scale parameters $r_1$, $r_2$, and the angle $\theta$ of the anisotropy. 

To reduce the number of parameters, we use the correlation function instead of the covariance function, which only depends on the ratio and not on the magnitude of the variances of streamfunction and velocity potential \citep {Daley1991}. This is possible as we can estimate the variance of the zonal and meridional wind with very low uncertainty due to the large size of the considered grid.

CL was maximized using the built-in function optim of \cite{R}. In order to show the independence of the optimization technique of the initial values it was started 50 times with varying initial parameters. This reveals that there is a single global maximum of the likelihood function.

Parameter uncertainty such as the Fisher information are not available for our problem. 
We thus resort to a parametric bootstrap \citep{Tibshirani1994} to assess uncertainty of the parameter estimates. We simulated the multivariate GRF using circulant embedding \citep{wood1994} to obtain independent realizations of the fitted process.
Re-estimating the parameters for a sample of 100 independent realizations provides the uncertainty of the parameter estimates given that the estimated model is true.
The simulation of the data was made possible by the implementation of the considered covariance model in an upcoming version of the spatial statistics package RandomFields \citep{RandomFields}.

\section{Results}\label{Results}

\begin{figure}[t]
  \centering  
\includegraphics*[width=1.11\textwidth]{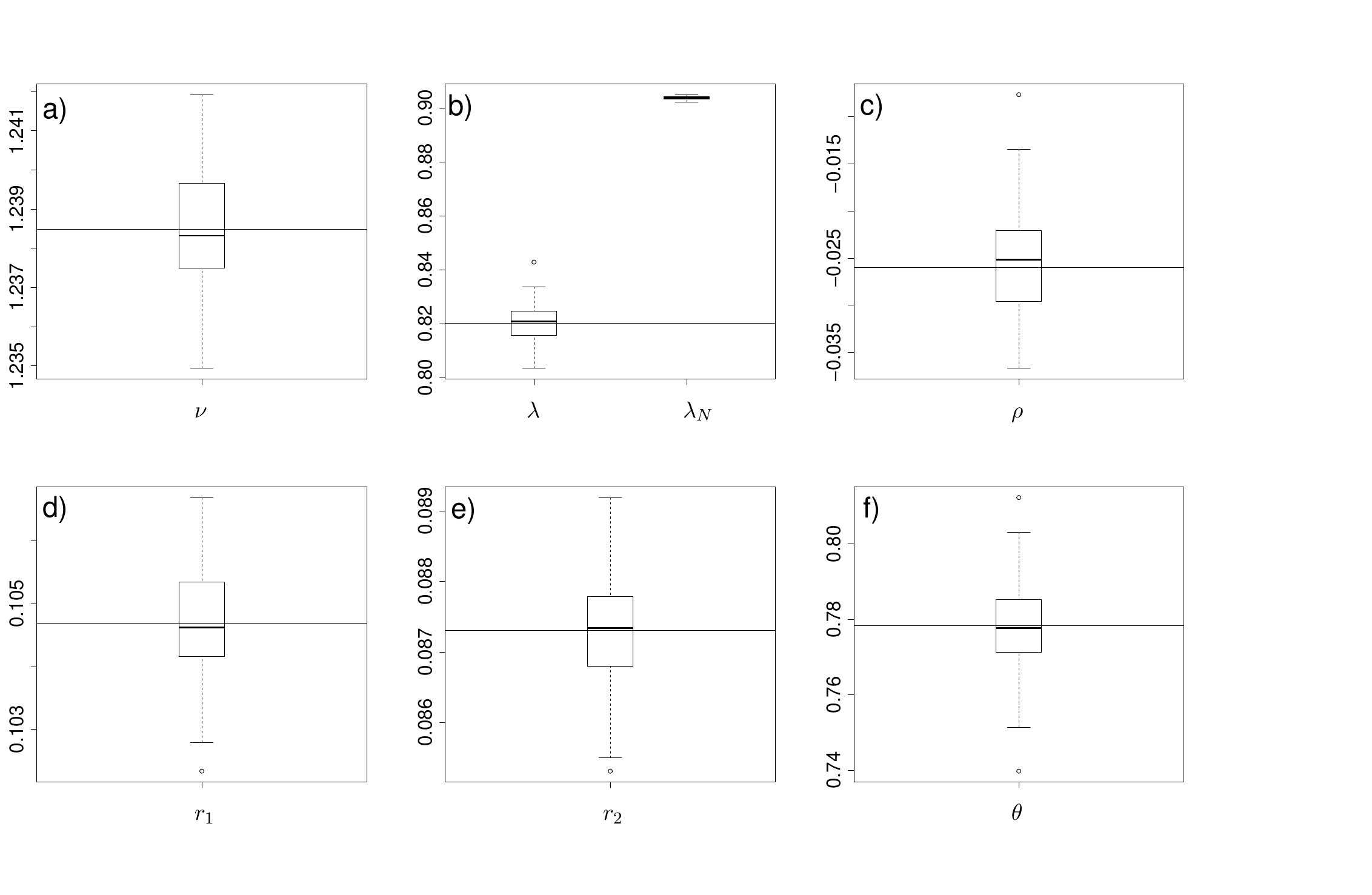}
  \caption {	Box-whisker plots representing the parametric bootstrap estimates for the inner-LBC wind anomalies. The horizontal lines indicate the ML estimates values: a) shows the smoothness parameter $\nu$, the left box-whisker in b) represents the ML estimates ($\lambda$), and the right box-whisker  the numerically derived estimates ($\lambda_N$) of the ratio $\lambda = \sigma_\chi/\sigma_\psi$. c) shows the correlation $\rho$,  d) the scale parameters $r_1$ e) the scale parameter $r_2$, and f) the angle of the anisotropy matrix $\theta$.}  
   \label{fig:bootstrap}
\end{figure}
Figure \ref {fig:bootstrap} shows the estimates of the parameters of the multivariate GRF and the respective distribution of the parametric bootstrap estimates as a boxplot. The ratio between divergent and rotational wind is estimated to about $\sigma_\chi/\sigma_\psi\approx 0.82$. This indicates, that both wind components are of the same order of magnitude. 
A geostrophic balance would require a ratio of order $\sigma_\chi/\sigma_\psi=0.1$, with a significant dominance of the rotational wind component. This is not the case in COSMO-DE, which is well consistent with the mesoscale dynamics, which are highly non-geostrophic. The results are also consistent with Bierdel (2012)\footnote{Personal communication: Lotte Beata Bierdel (2012): Mesoskalige Turbulenz in dem konvektionsaufl\"osenden Wettervorhersagemodell {COSMO-DE-EPS}. Masterarbeit in Meteorologie. Meteorologisches Institut der Friedrich-Wilhelms-Universit\"at Bonn. 159p.}. Her spectral analysis of the horizontal wind fields of COSMO-DE-EPS revealed a slightly stronger rotational than divergent component. Figure \ref {fig:bootstrap}b compares the statistical estimate for $\lambda=\sigma_\chi/\sigma_\psi$ to a numeric estimate, which equals the ratio of the $L^2$-norms of curl and divergence of the wind field calculated with finite difference approximations and which is denoted by $\lambda_N$. 
Since the parametric bootstrap is performed on simulated data, we know the true values corresponding to the data. This allows to compare different estimates of ratio of divergent and rotational variance.
Though the statistical estimate has a higher variance it clearly outperforms the numeric estimate due to the relatively large bias of the latter. Our methods provide a possibility to test numeric estimates for their consistency.

The correlation between streamfunction and velocity potential $\rho$ is almost zero  $\approx -2\times 10^{-2}$. Similar results have been described for larger scales \citep{Hollingsworth1986} and have often been assumed in the literature \citep {Daley1991}.
The smoothness parameter $\nu$ is close to $1.24$. This corresponds to non-continuous fields of vorticity and divergence. This relatively low value of $\nu$ is not due to noise in the data. We have included  	tentatively a noise parameter in the estimation but it was set to zero.
As a measure for the anisotropy we consider the ratio of the scale parameters $r_1/r_2$. This ratio is significant larger than 1 for both data sets, which shows that the data is anisotropic. 
The estimated parameters are very much in accordance with our expectations, as they describe a non-geostrophic and anisotropic wind field. The most important result is that the independence  of streamfunction and velocity potential in the case of the 5 June 2011 is valid on the mesoscale. Similar results were already known for larger scales \citep {Hollingsworth1986}. In addition, our parametric bootstrap reveals that this covariance model can be estimated with a very high precision on model data. We  have shown that our estimate of the ratio of divergence and vorticity is superior to a numeric estimate on data simulated by our model.

\begin{figure}
 \includegraphics*[width=\textwidth,page=1]{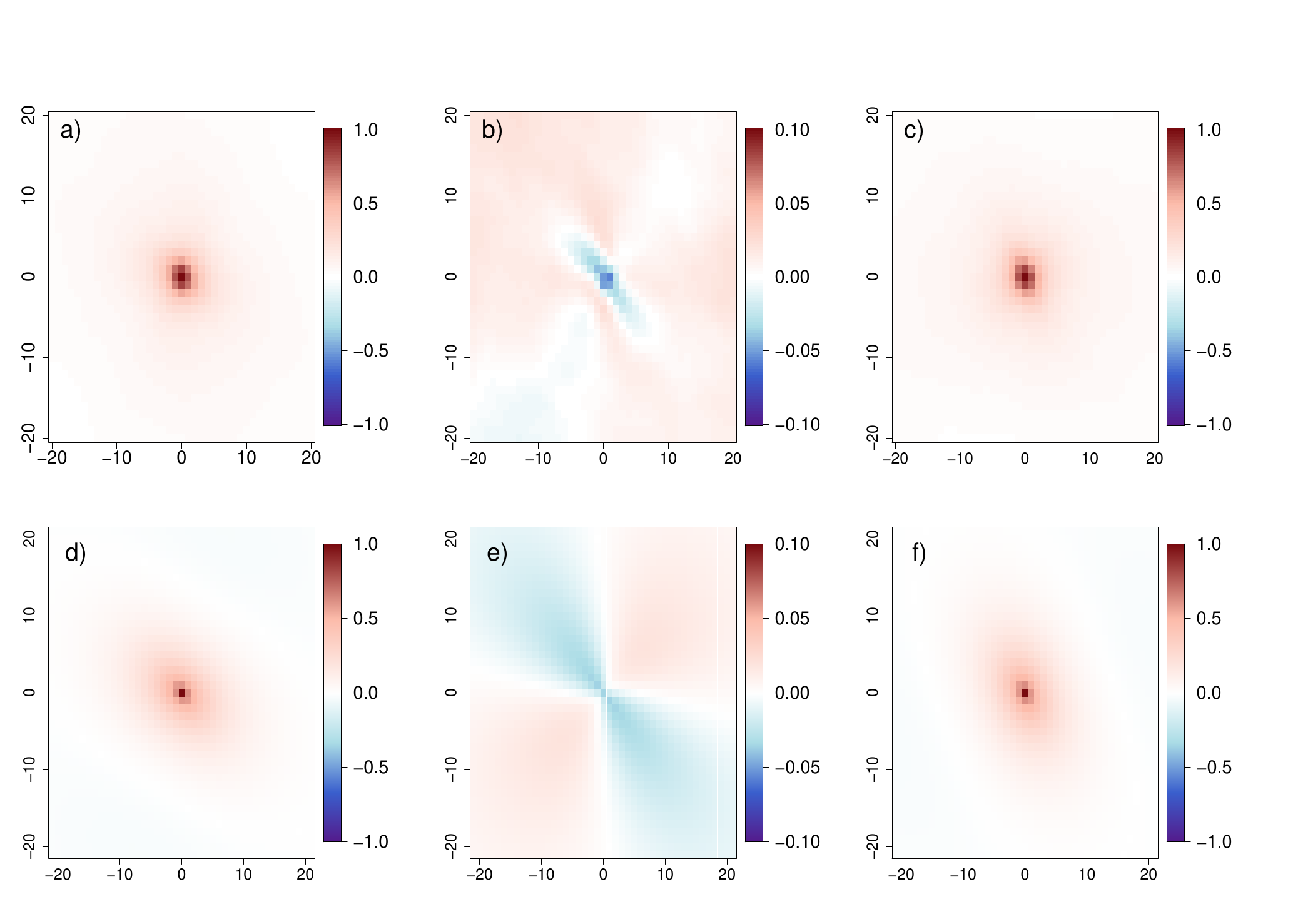} 
 \caption {Empirical correlation (above) and estimated correlation (below) for data set 1.  a) (u,u) empirical correlation; 
 b) (u,v) empirical correlation; c) (v,v) empirical correlation; d) (u,u) estimated correlation; e) (u,v) estimated correlation; f) (v,v) estimated correlation.}
 \label {physical_cov}
\end{figure}

Fig. \ref {physical_cov} shows the empirical estimate of the  correlation structure of the data and the correlation obtained for the maximum likelihood estimation. Again the scale and the orientation of the correlation is very well matched. The $(u,u)$ and $(v,v)$ auto correlation component is matched relatively well. The $(u,v)$ correlation component has a deviation from the data as there are regions of positive correlation, which is not present in the empirical correlation estimate.

 The implementation of our covariance model in an upcoming version of the R package RandomFields \citep{RandomFields} allows for the simulation of large field with a size of the order of $\left( 800\times800\right)$ grid points. This is made feasible by using circulant embedding introduced by \cite{wood1994}. Circulant embedding  is a powerful simulation technique, which to the best of our knowledge, has not been used for the simulation of wind fields yet. 
 
\begin{figure}
\centering
 \includegraphics[width=\textwidth,keepaspectratio=TRUE]{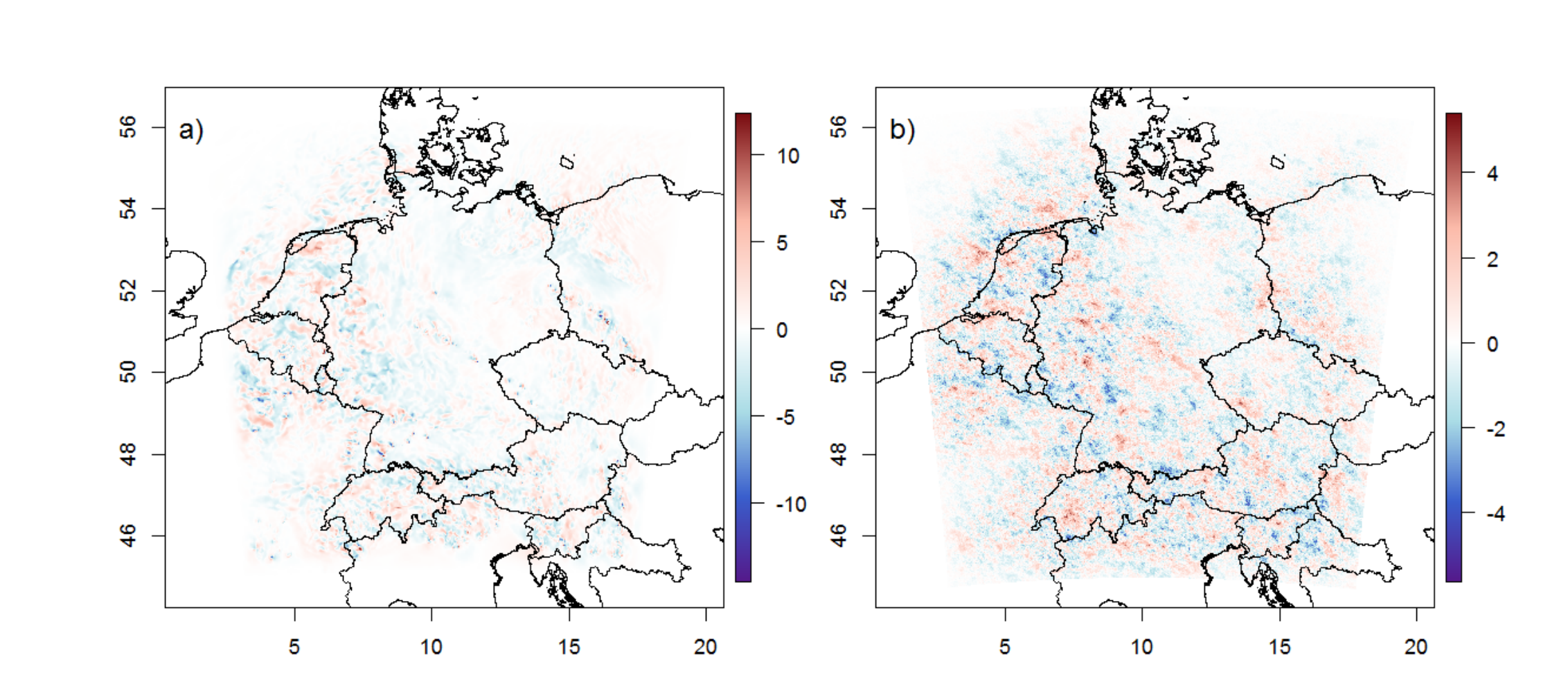} 
  \caption {	a) Same as Figure \ref {fig:simuset1}b). b) Zonal wind component of a realization of the fitted GRF. The x/y- axis are in longitude and latitude. }
    \label {fig:simuset2}
  \end{figure}

 Figure \ref{fig:simuset2} shows the zonal wind anomalies from Fig.\ \ref{fig:simuset1} together with a realization of the fitted multivariate GRF, which has been scaled with the spatial variance that has not been resolved by the transformation \eqref {transform}. It shows that the orientation as well as the spatial scale of the zonal wind fields match very well. The multivariate GRF shows less extreme  values and less values very close to zero, due to the assumption of Gaussianity.
However, visual accordance is quite well, such that we conclude that the multivariate GRF formulation represents a useful stationary, multivariate  Gaussian random fields approximation of mesoscale wind anomalies.

\section {Conclusions}\label {Conclusions}
In this paper we introduce a multivariate GRF which jointly models  streamfunction, velocity potential, the 2-dimensional wind field, vorticity and divergence. Its flexibility allows for different variances of the potential functions, anisotropy and a flexible smoothness parameter. Further, the model is able to represent non-zero correlation of the divergent and non-divergent wind component.  All parameters of the proposed covariance model have direct meteorological interpretation, such that they  provide meteorological insight into the dynamics of the atmosphere. Further, the model allows us to easily implement meteorological balances such as non-divergence or geostrophy.
 
We have reviewed the theory that guarantees the existence of derivatives of stochastic processes, developed a complex covariance model for various atmospheric variables and studied its transformation subject to anisotropy. 
Our multivariate GRF is a counter example to a theorem of \cite{obukhov1954statistical}, which claims that the rotational and divergent components of an isotropic vector field are necessarily uncorrelated.

We have developed an estimation technique and shown its performance for wind anomalies of a mesoscale ensemble prediction system (COSMO-DE-EPS). A parametric bootstrap method provides  estimates of the uncertainty implicit in our estimation technique. We thus provide estimates for the ratio of variances of the rotational and divergent wind component without numerical approximations. Numeric estimates suffer from a truncation error, which arises due to the numerical scheme that computes the derivatives of the wind field. 

The multivariate GRF formulation may be particularly useful for global atmospheric models with a spectral representation of the horizontal fields, such as the ECHAM climate model \citep{Roeckner03}. 
Spectral models solve the prognostic equations for the potentials instead of the horizontal wind components, whereas the observations are given as horizontal wind vectors. Our multivariate GRF formulation provides a consistent formulation of the covariance structure for both the potential and the horizontal wind components. A stochastic formulation of the potentials may also be relevant for the assimilation of  measurements of the vertical velocity \citep {buhl2015}, which provide proxies for the horizontal divergence of the field. Our covariance function represents the divergence within a stochastic model, which is needed to assimilate the observations.

The proposed covariance model can be used to interpolate observed wind fields and to compute the associated derivative fields. This is feasible either by conditional simulation or Kriging. Numerical methods have been used for interpolation  \citep{Schaefer1979} and the computation of derivatives of vector fields \citep {Caracena1987, doswell1988}.  While numeric methods become significantly more complex for scattered observations, the multivariate GRF formulation provides an accessible way for both problems which additionally provides information about the uncertainty.
If  \textit {for example} the expected value of streamfunction and vector potential given a certain wind field is computed, this approach can be considered as a stochastic model for the solution of partial differential equations. As stochastic models describe the uncertainty of all of the variables these methods even allow stochastic error bands to be computed for the solution of the partial differential equations.

Another potential application is the stochastic simulation of the transport of tracer variables such as aerosols or humidity in the atmosphere. 
Stochastic models that describe gradient fields and their divergence have been considered in the literature \citep{Scheuerer2012}. However, to the best of our knowledge no stochastic model has been formulated to jointly model spatial  wind fields and its divergence. Both variables are needed to describe the transport adequately. 

Our methods show that both physical coherence and geostrophic constraints can be easily implemented into a covariance model. Further, we have illustrated that the model parameters can be estimated with very small uncertainty.
Using kriging techniques our methods allow to compute mean streamfunction and mean velocity potential for a given wind field. This  can be considered as a stochastically motivated solution of partial differential equations.

%
\acknowledgments
R\"udiger Hewer was funded by VolkswagenStiftung
within the project ''Mesoscale Weather Extremes Theory --
Spatial Modeling and Prediction (WEX-MOP)''. Data used
in this study are kindly provided by the German Meteorological
Service (DWD). We thank Chris Snyder and an anonymous reviewer for the thoughtful comments, that improved our paper substantially. Especially we are grateful to the reviewer for the idea to transform the data such that our model assumptions are more appropriate.
We thank Sebastian Buschow for help in preparing the data.

%






%
%
%
\appendix[A]
\appendixtitle {Positive definiteness of Daley's (1985) model}
\label {Daley}
 \cite{Daley1985} proposed the covariance model \citep [cf.][]{moreva2016,Gneiting2010a}

\begin{align*} 
 C \left( r\right)=\left(\begin{matrix}
\exp\left( {-\frac{1}{2}r^2}\right) & \lambda \exp \left( -\frac{1}{2}\left(\frac{r}{a}\right)^2\right)  \\
\lambda \exp \left( -\frac{1}{2}\left(\frac{r}{a}\right)^2\right) & \exp \left( -\frac{1}{2}\left(\frac{r}{a}\right)^2\right)
\end{matrix}\right), \quad r=\|h\|,
 \end{align*}
 for streamfunction and velocity potential. More general covariance models ofth
The Fourier transform of this covariance matrix is given by
\begin{align*} 
 \mathfrak {F} \left( C\right)\left( \varphi\right)=\left(\begin{matrix}
\exp \left( -\frac{1}{2}\varphi^2\right) & \frac{\lambda}{a} \exp \left( -\frac{1}{2}\left( \frac{\varphi}{a}\right)^2\right) \\
 \frac{\lambda}{a} \exp \left( -\frac{1}{2}\left( \frac{\varphi}{a}\right)^2\right) &  \frac{1}{a} \exp \left( -\frac{1}{2}\left( \frac{\varphi}{a}\right)^2\right)
\end{matrix}\right).
 \end{align*}
By Cram\'er Theorem \citep{chiles2009geostatistics} this Fourier-transform needs to be positive definite for almost all frequencies $\varphi$. This is equivalent to
\begin{align*} 
\det \left(  \mathfrak {F} \left( C\right)\left( \varphi\right)\right)\ge 0 \quad \forall \varphi \in \R
 \end{align*}
a condition equivalent to
\begin{align*} 
\exp \left( -\frac{1}{2}\varphi^2 \left( 1-\frac{1}{a^2}\right)\right)\ge \frac{\lambda^2}{a}\quad \forall \varphi \in \R. \
 \end{align*}
 If $a>1$ the model is not positive definite unless $\lambda=0$.
 If $0<a\le1$ the model is positive definite if $a\ge \lambda^2$.
 Daley proposed $a>1$  such that the model does not allow a non-zero correlation.

 \appendix [B]
 \appendixtitle {Obukhov's (1954) independence claims}
\cite{obukhov1954statistical} presents two arguments for an isotropic rotational  field having zero  correlation with an isotropic scalar field and with an isotropic gradient field. We believe that both arguments are erroneous for the same reason. As the argument for the scalar field is much less involved, we restrict ourselves to this case.
 Obukhovs claims that the covariance of an isotropic rotational field to an arbitrary scalar isotropic variable is of the form
 $\Cov \left( \chi_s,\rot \psi_{s+h}\right)= P \left( \|h\|\right) h/\|h\|$
 for some function $P$. 
Using the non-divergence of a rotational field Obukhov deduces from his assumption:
\begin{align*} 
 &0=\E \left( \chi_s \div\rot \psi_{s+h}\right)=\div  \E \left( \chi_s \rot \psi_{s+h}\right)=\div \left(\frac{P \left( \|h\|\right)}{\|h\|}\left(\begin{array}{c} h_1 \\ h_2 \end{array}\right)\right)\\
 &= \frac{2P \left( \|h\|\right)}{\|h\|}+\frac{\partial}{\partial \|h\|}\left( \frac{P \left( \|h\|\right)}{\|h\|}\right)\|h\|
 =\frac{P \left( \|h\|\right)}{\|h\|}+P^\prime \left( \|h\|\right)
 \end{align*} 
This differential equation is solved by the function 
\begin{align*} 
 P \left( \|h\|\right)=\frac{c}{\|h\|} \quad \quad c\in \R.
 \end {align*}
 If $c\neq 0$ this function has a pole. This implies that the variance of the corresponding field could not exist. Hence $c=0$ and this again implies the zero correlation between the scalar field and the rotational field. 
 
We believe that the correct covariance of a scalar field and a rotational field is given by
 \begin{align*} 
   \Cov \left( \chi_s,\rot \psi_{s+h}\right)= \frac{P \left( \|h\|\right)}{\|h\|}\left(\begin{array}{c} -h_2 \\ h_1 \end{array}\right),
 \end{align*}
 for some P, as the curl operator derives the first component in direction $e_2$ and the second in direction $e_1$. This covariance is consistent with the anisotropic transformation of the field, which has been described in \eqref {rotational}. Using this assumption the independence of an isotropic scalar field and an isotropic rotational field cannot be deduced. However,
 \begin{align*} 
 &\E \left( \chi_s \div\rot \psi_{s+h}\right)=\div  \E \left( \chi_s \rot \psi_{s+h}\right)=\div \left(\frac{P \left( \|h\|\right)}{\|h\|}\left(\begin{array}{c} -h_2 \\ h_1 \end{array}\right)\right)\\
 &=\frac{P \left( \|h\|\right)}{\|h\|^2}\left( -h_2h_1+h_1h_2\right)=0
 \end{align*} 
for any differentiable function $P$.
\appendix[C]
\appendixtitle {Formulae of the isotropic covariance model}
We describe the formula for the covariance function considered in this paper (equation \eqref {full_covariance}). For brevity we introduce the following notation
\begin{align*}  
 & X_{1,s}\defeq \psi_s
 & X_{2,s}\defeq \chi_s\\
 & U_{1,s}\defeq u_s
 & U_{2,s}\defeq v_s\\
 & \partial_i\defeq \partial_{e_i},
\end {align*}
  and omit the argument of the covariance functions, which is $\left( t-s\right)$ in all of the following cases.
\begin{align*} 
	\begin {split}
	& C^{i,j}\defeq\Cov \left( X_{i,s},X_{j,t}\right)\\
	& \Cov \left( U_{i,s},U_{j,t}\right)=
	\left( \minus1\right)^{i+j}\frac{\partial^2}{\partial_{3-i}\partial_{3-j}}C^{1,1}+
	\left( \minus1\right)^{i}\frac{\partial}{\partial_{3-i}\partial_{j}}C^{1,2}+\left( \minus1\right)^{j}\frac{\partial^2}{\partial_{i}\partial_{3-j}}C^{2,1}+\frac{\partial}{\partial_i\partial_j}C^{2,2}\\
	& \Cov \left( \Delta X_{i,s},\Delta X_{j,t}\right)=\sum_{\left( k,l\right)\in \{1,2\}^2}\frac{\partial^4}{\partial^2_k\partial^2_l}C^{i,j}\\
	&\Cov \left( U_{i,s},X_{j,t}\right)=- \Cov \left( X_{j,s},U_{i,t}\right)=\left(\minus1\right)^{i}\frac{\partial}{\partial e_{3-i}}C^{1,j}+\frac{\partial}{\partial e_i}C^{2,j}\\
	&\Cov \left( X_{i,s},\Delta X_{j,t}\right)=\Cov \left( \Delta X_{j,s},X_{i,t}\right)=\frac{\partial^2}{\partial^2 e_1}C^{i,j}+\frac{\partial^2}{\partial^2 e_2}C^{i,j}\\
	&\Cov \left( U_{i,s},\Delta X_{j,t}\right)=-\Cov \left( \Delta X_{j,s},U_{i,t}\right)\\
	&\quad=\left( \minus1\right)^i \frac{\partial^3}{\partial_{3-i}\partial^2_{1}}C^{1,j}+\left( \minus1\right)^i\frac{\partial^3}{\partial_{3-i}\partial^2_{2}}C^{1,j}+\frac{\partial^3}{\partial_i\partial^2_1}C^{2,j}+\frac{\partial^3}{\partial_i\partial^2_{2}
	}C^{2,j},
	\end {split} 
 \end{align*}
where $i,j\in \{1,2\}$ .

 \bibliographystyle{ametsoc2014}
 \bibliography{bib}

%

%

\end{document}